\documentstyle[graphics]{elsart}
\begin{document}
\begin{frontmatter}
\title{Metal-Insulator transitions in generalized Hubbard models}
\thanks[contact]{ Corresponding author: Erik Koch,
  Max-Planck-Institut f\"ur Festk\"orperforschung,
  Heisenbergstra\ss e 1, 70569 Stuttgart, Germany;
  phone: +49 711 689 1666, fax:  +49 711 689 1632,
  email koch@and.mpi-stuttgart.mpg.de }
\author{Erik Koch, Olle Gunnarsson}
\address{Max-Planck-Institut f\"ur Festk\"orperforschung, 
         70569 Stuttgart, Germany}
\author{Richard M. Martin}
\address{Department of Physics, University of Illinois, 
         Urbana, IL 61801, USA}

\begin{abstract}
We study the Mott transition in Hubbard models with a degenerate band on
different 3-dimensional lattices. While for a non-degenerate band only the 
half-filled system may exhibit a Mott transition, with degeneracy there 
can be a transition for any integer filling. We analyze the filling dependence
of the Mott transition and find that $U_c$ (the Hubbard interaction $U$ at 
which the transition takes place) decreases away from half-filling.
In addition we can change the lattice-structure of the model. This allows us
to study the influence of frustration on the Mott transition. We find that
frustration increases $U_c$, compared to bipartite systems.
The results were obtained from fixed-node diffusion Monte Carlo calculations
using trial functions which allow us to systematically vary the magnetic 
character of the system. 
To gain a qualitative understanding of the results, we have developed simple 
hopping arguments that help to rationalize the doping dependence and the 
influence of frustration on the Mott transition.
Choosing the model parameters to describe the doped Fullerides, we can make
contact with experiment and understand why some of the Fullerides are metals,
while others, which according to density functional theory should also be 
metallic, actually are insulators.
\end{abstract}
\begin{keyword}
 Mott transition, Hubbard model, degenerate, filling,
 bipartite, frustration, Fullerenes, quantum Monte Carlo
\end{keyword}
\end{frontmatter}

\section{Introduction}

The Hubbard model is an elementary model to study strongly interacting 
systems. In particular it can be used to understand the Mott metal-insulator
transition. The original Hubbard model describes $s$ electrons in a narrow
band. With one electron per site, the half-filled system will become insulating
for large enough correlation strength. To describe more
realistic situations one has to generalize the Hubbard model to systems with
degenerate (or near-degenerate) orbitals. Such degenerate orbitals can arise 
for example in molecular solids or transition metal compounds. For such a
degenerate Hubbard model there will be a Mott transition not only at 
half-filling, but for all {\em integer fillings}. It is then a natural question
how the location of the Mott transition depends, for an otherwise unchanged
Hamiltonian, on the filling. An other issue is the problem of how the
Mott transition depends, for the same filling, on the lattice-structure of
the system. 
To address these questions we have determined the Mott transition for 
degenerate Hubbard models with various integer fillings and different
lattice-structures. We have used Hamiltonians that describe the alkali doped
Fullerenes, since these materials have been synthesized in various integer 
dopings and crystal structures.

\section{Model and Method}

The inter-molecular interaction in solid C$_{60}$ is very weak.
Therefore the energy levels of the molecule merely broaden into narrow, 
well separated bands \cite{ldabands}. The conduction band originates from 
the lowest unoccupied molecular orbital, the 3-fold degenerate $t_{1u}$ 
orbital. To get a realistic, yet simple description of the electrons in the
$t_{1u}$ band, we use a Hubbard-like model that describes the interplay
between the hopping of the electrons and their mutual Coulomb
repulsion \cite{c60mott}:
\begin{equation}\label{Hamil}
H=\sum_{\langle ij\rangle} \sum_{mm'\sigma} t_{im,jm'}\;
              c^\dagger_{im\sigma} c^{\phantom{\dagger}}_{jm'\sigma}
 +\;U\sum_i\hspace{-0.5ex} \sum_{(m\sigma)<(m'\sigma')}\hspace{-1ex}
       n_{i m\sigma} n_{i m'\sigma'} .
\end{equation}
The sum $\langle ij \rangle$ is over nearest-neighbor sites.
The hopping matrix elements $t_{im,jm'}$ between orbital $m$ on molecule $i$
and orbital $m'$ on molecule $j$ are obtained from a tight-binding
parameterization \cite{TBparam,A4C60}. The molecules are orientationally
disordered \cite{oridisord}, and the hopping integrals are chosen such that
this orientational disorder is included \cite{hopdisord}. The band-width for
the infinite system is $W=0.63\,eV$. The on-site Coulomb interaction is
$U\approx 1.2\,eV$.
The model neglects multiplet effects, but we remark that these tend to be
counteracted by the Jahn-Teller effect, which is also not included in the
model.

To identify the Mott transition we calculate the energy gap
\begin{equation}\label{gap}
  E_g=E(N+1)-2\,E(N)+E(N-1) ,
\end{equation}
where $E(N)$ is the energy of a cluster of $N_{\rm mol}$ molecules with
$N=n\,N_{\rm mol}$ electrons (integer filling $n$). We determine these 
energies using fixed-node diffusion Monte Carlo \cite{DMC}. Starting from a
trial function $|\Psi_T\rangle$ we calculate
\begin{equation}\label{proj}
  |\Psi^{(n)}\rangle = [1-\tau(H-w)]^n\;|\Psi_T\rangle ,
\end{equation}
where $w$ is an estimate of the ground-state energy. The $|\Psi^{(n)}\rangle$
are guaranteed to converge to the ground state $|\Psi_0\rangle$ of $H$, if
$\tau$ is sufficiently small and $|\Psi_T\rangle$ is not orthogonal to
$|\Psi_0\rangle$. Since we are dealing with Fermions, the Monte Carlo
realization of the projection (\ref{proj}) suffers from the sign-problem.
To avoid the exponential decay of the signal-to-noise ratio we use
the fixed-node approximation \cite{DMC}. For lattice models this involves
defining an effective Hamiltonian $H_{\rm eff}$ by deleting from $H$ all
nondiagonal terms that would introduce a sign-flip. Thus, by construction,
$H_{\rm eff}$ is free of the sign-problem. To ensure that the ground-state
energy of $H_{\rm eff}$ is an upper bound of $E_0$, for each deleted hopping
term an on-site energy is added in the diagonal of $H_{\rm eff}$. Since
$|\Psi_T\rangle$ is used for importance sampling, $H_{\rm eff}$ will depend
on the trial function. Thus, in a fixed-node diffusion Monte Carlo calculation 
for a
lattice Hamiltonian we choose a trial function and construct the corresponding
effective Hamiltonian, for which the ground-state energy $E_{\rm FNDMC}$ can
then be determined without sign-problem by diffusion Monte Carlo.

\begin{figure}
\centerline{\resizebox{7cm}{!}{\includegraphics{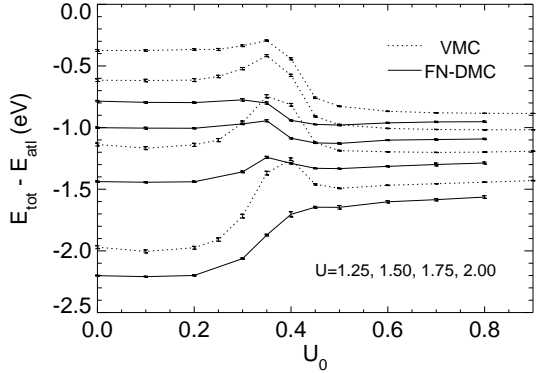}}}
\caption{\label{corrU0}
          Dependence of variational (VMC) and fixed-node diffusion
          Monte Carlo (FN-DMC) on the trial function. $U_0$ is the Hubbard
          interaction that was used for the Slater determinant in the
          Gutzwiller wavefunction $\Psi_T(R)=g^{D(R)}\;\Phi(U_0)$.
          The Gutzwiller parameter $g$ has always been optimized.
          The results shown here are the energies (relative to the atomic
          limit) for a Hamiltonian that describes K$_3$C$_{60}$ (32 sites),
          with $U$ being varied from $1.25$ (lowest curve) to $2.00\,eV$
          (highest curve).}
\end{figure}

For the trial function we make the Gutzwiller Ansatz
\begin{equation}\label{psitrial}
  |\Psi(U_0,g)\rangle = g^D\;|\Phi(U_0)\rangle ,
\end{equation}
where the Gutzwiller factor reflects the Coulomb term
$U\,D=U\sum n_{i m\sigma} n_{i m'\sigma'}$ in the Hamiltonian (\ref{Hamil}).
$|\Phi(U_0)\rangle$ is a Slater determinant that is constructed by solving
the Hamiltonian in the Hartree-Fock approximation, replacing $U$ by a
variational parameter $U_0$. Increasing $U_0$ will change the character of 
the trial function from paramagnetic to antiferromagnetic. This transition 
is also reflected in the variational energies obtained in quantum Monte Carlo,
as shown in Fig.~\ref{corrU0}. Clearly, for small $U$ the paramagnetic
state is more favorable, while for large $U$ the antiferromagnetic state 
gives a lower variational energy. Details on the character of the trial 
function and the optimization of the parameters can be found in 
Ref.\ \cite{corrsmpl}.

\section{Results}

We now turn to the problem of the Mott transition in degenerate Hubbard
models for different integer dopings and for different lattice-structures.
The examples will be for integer-doped Fullerides A$_n$C$_{60}$, where
A stands for an alkali metal like K, Rb, or Cs.
Density functional calculations predict that all the doped Fullerides 
A$_n$C$_{60}$ with $n=1\ldots5$ are metals \cite{ldabands}. Only
C$_{60}$ and A$_6$C$_{60}$ are insulators with a completely empty/filled 
$t_{1u}$ band. On the other hand, Hartree-Fock calculations for the
Hamiltonian (\ref{Hamil}) predict a Mott transition already for
$U$ smaller than the band-width, and hardly any doping dependence.
General arguments also suggest that the alkali doped Fullerenes should
be Mott insulators, since the Coulomb repulsion $U$ between two electrons
on the same C$_{60}$ molecule ($U\approx 1.2\,eV$) is substantially larger 
than the width of the $t_{1u}$ band ($W\approx 0.6\,eV$). It has therefore
even been suggested that experimental samples of, say, the superconductor
K$_3$C$_{60}$ are metallic only because they are non-stoichiometric, 
i.e.\ that they actually are K$_{3-\delta}$C$_{60}$ \cite{lof}.

\subsection{K$_3$C$_{60}$}

\begin{figure}
  \centerline{\resizebox{3.3cm}{!}{\includegraphics{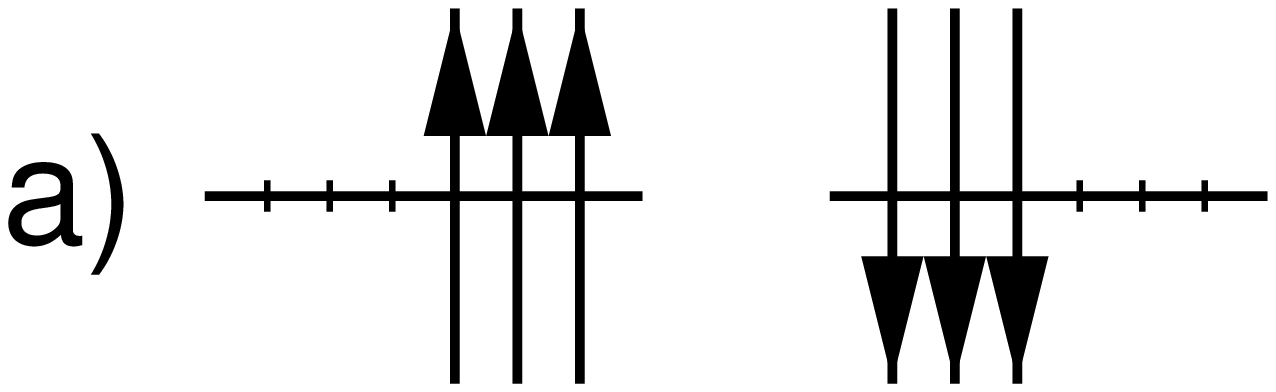}}}

  \vspace{0.5ex}
  \centerline{\resizebox{3.3cm}{!}{\includegraphics{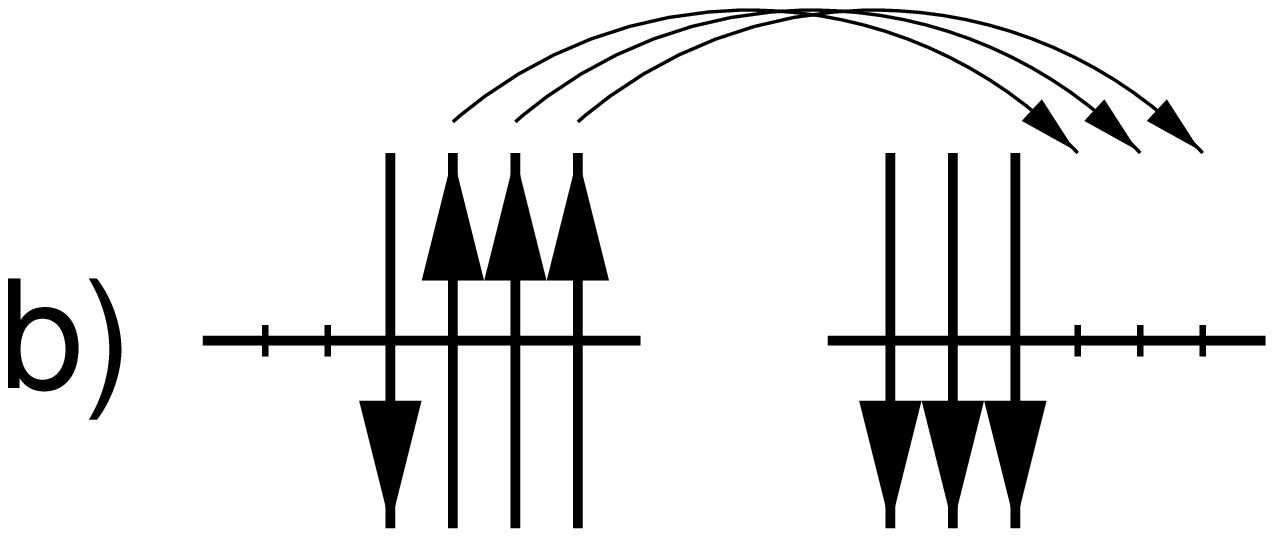}}}
  \caption[]{\label{hopping}
             Degeneracy argument:
             a) In the large-$U$ limit hopping is suppressed for an 
                integer-filled system.
             b) An additional charge can hop without extra Coulomb energy.
                For degenerate orbitals and an antiferromagnetic background
                there are several different ways the extra charge can hop
                to a neighboring molecule.
            }
\end{figure}

In a first step we investigate what consequences the degeneracy of the 
$t_{1u}$-band has for the Mott transition in K$_3$C$_{60}$. The analysis
is motivated by the following simple argument \cite{c60mott,degen}. In the 
limit of very large $U$ we can estimate the energies needed to calculate the 
gap (\ref{gap}). For half-filling, all molecules will have three electrons in 
the $t_{1u}$ orbital (Fig.~\ref{hopping} a). Hopping is strongly suppressed 
since it would increase the energy by $U$. Therefore, to leading order in 
$t^2/U$, there will be no kinetic contribution to the total energy $E(N)$. 
In contrast, the systems with $N\pm1$ electrons have an extra electron/hole 
that can hop without additional cost in Coulomb energy. To estimate the kinetic
energy we calculate the matrix element for the hopping of the extra charge 
against an antiferromagnetic background. Denoting the initial state with extra 
charge on molecule $i$ by $|1\rangle$, we find that the second moment
$\langle1|H^2|1\rangle$ is given by the number of different possibilities 
for a next-neighbor hop times the single electron hopping matrix element $t$
squared. By inserting $\sum_j |j\rangle\langle j|$, where 
$|j\rangle$ denotes the state with the extra charge hopped from site $i$ to 
site $j$, we find $\langle1|H|j\rangle = \sqrt{3}\,t$, since, with an 
antiferromagnetic background and degeneracy 3, there are three different ways 
an extra charge can hop to a neighboring molecule (Fig.~\ref{hopping}, b). Thus,
due to the 3-fold degeneracy, the hopping matrix element is enhanced by a 
factor $\sqrt{3}$ compared to the single electron hopping matrix element $t$.
In the single-electron problem the kinetic energy is of the order of half
the band-width $W/2$. The enhancement of the hopping matrix element in the
many-body case therefore suggests that the kinetic energy for the extra charge
is correspondingly enhanced. Inserting the energies into (\ref{gap}) 
we find that for the 3-fold degenerate system our simple argument predicts 
a gap 
\begin{equation}\label{Egap}
  E_g=U-\sqrt{3}\,W , 
\end{equation}
instead of $E_g=U-W$ in the non-degenerate case. 
Extrapolating to intermediate $U,\,$ it appears that the degeneracy 
shifts the Mott transition towards larger $U$.

The above argument is, of course, not rigorous. First, it is not clear 
whether the result for $E_g$ that was obtained in the limit of large $U$ 
can be extrapolated to intermediate $U,\,$ where the Mott transition 
actually takes place. Also the analogy of the hopping in the many-body case 
with the hopping of a single electron is not rigorous, since the hopping of an
extra charge against an antiferromagnetic background creates a string of
flipped spins \cite{Nagaoka}. Nevertheless the argument suggests that orbital
degeneracy should play an important role for the Mott transition.

\begin{figure}
 \centerline{\resizebox{8cm}{!}{\includegraphics{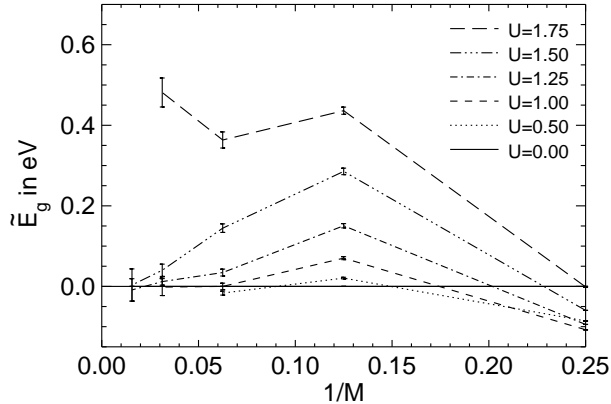}}}
 \caption[]{\label{Egap3}
          Finite-size corrected gap $\tilde{E}_g=E_g-U/M-E_g(U=0)$ for
          increasing Coulomb interaction $U$ as a function of $1/M$, 
          where $M$ is the number of molecules. The calculations are for
          a Hubbard model with hopping matrix elements appropriate for
          K$_3$C$_{60}$. The band-width varies between $W=0.58\,eV$ for
          $M=4$ and $W=0.63\,eV$ in the infinite-size limit.} 
\end{figure}

To check this proposition, we look at the results of the quantum Monte Carlo 
calculations for the model Hamiltonian (\ref{Hamil}) \cite{c60mott}. The 
Coulomb interaction $U$ has been varied from $U=0\ldots 1.75\,eV$ to study 
the opening of the gap. Since the Monte Carlo calculations are for finite 
systems, we have to extrapolate to infinite system size. 
To improve the extrapolation we correct for finite-size effects: First,
there could be a gap $E_g(U=0)$ already in the spectrum of the non-interacting
system. Further, even for a metallic system of $M$ molecules, there will be a 
finite-size contribution of $U/M$ to the gap. It comes from the electrostatic 
energy of the extra charge, uniformly distributed over all sites. Both 
corrections vanish in the limit $M\to\infty$, as they should. The finite-size 
corrected gap $\tilde{E}_g=E_g - U/M - E_g(U=0)$ for systems with 
$M=$ 4, 8, 16, 32, and 64 molecules is shown in Fig.~\ref{Egap3}. We find that 
the gap opens for $U$ between $1.50\,eV$ and $1.75\,eV.\,$ Since for the real 
system $U=1.2\ldots1.4\,eV,\,$ K$_3$C$_{60}$ is thus close to a Mott 
transition, but still on the metallic side --- even though $U$ is considerably
larger than the band-width $W$. This is in contrast to simpler theories that
neglect orbital degeneracy.

\subsection{Doping dependence}

\begin{figure}
 \centerline{\resizebox{7cm}{!}{\includegraphics{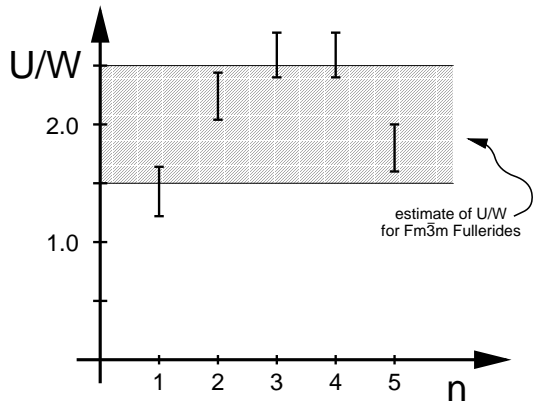}}}
 \caption[]{\label{Mott}
            Doping dependence of the Mott transition. The error bars indicate 
            the estimate of the critical ratio $U_c/W$ for different integer 
            fillings of the $t_{1u}$ band. The calculations are for 
            Fm${\bar 3}$m Fullerides with fcc lattice-structure and 
            orientational disorder. The shaded region shows the range
            of $U/W$ in which the doped Fullerenes are falling.}
\end{figure}

The degeneracy argument described above for K$_3$C$_{60}$ can be generalized 
to integer fillings. Away from half-filling the enhancement of the hopping 
matrix elements for an extra electron is different from that for an extra 
hole. The effective enhancement for different fillings are given in Table
\ref{enhance}. 
We find that the enhancement decreases as we move away from half-filling.
Therefore we expect that away from half-filling correlations become 
more important, putting the system closer to the Mott transition, or maybe
even pushing it across the transition, making it an insulator. We have 
analyzed the doping dependence of the Mott transition for the same
Hamiltonian as used for K$_3$C$_{60}$, changing the filling of the $t_{1u}$ 
band from $n=1$ to 5. This model describes the 
Fm${\bar 3}$m-Fullerides A$_n$C$_{60}$ with fcc lattice and orientational
disorder \cite{rmp}. The critical Coulomb interaction $U_c$, at which, for the
different integer fillings, the transition from a metal (for $U<U_c$) to an 
insulator ($U>U_c$) takes place, is shown in Fig.~\ref{Mott}. As expected from
the degeneracy argument, $U_c$ decreases away from $n=3$. 
We note, however, that $U_c$ is asymmetric around half-filling. 
This asymmetry is not present in the simple degeneracy argument, where we
implicitly assumed that the lattice is bipartite. In such a 
situation we have electron-hole symmetry, which implies symmetry around 
half-filling. For frustrated lattices, like the fcc lattice, electron-hole 
symmetry is broken, leading to an asymmetry in $U_c$ that is seen in 
Fig.~\ref{Mott}. 

\begin{table}
 \begin{centering}
 \begin{tabular}{l@{\hspace{5ex}}c@{$\;\approx\,$}c}
  \hline\hline
  filling & \multicolumn{2}{c}{enhancement}\\
  \hline
  $n=\;3$ & $\sqrt{3}$                  & 1.73\\[0.5ex]
  $n=2,4$ & ${\sqrt{3}+\sqrt{2}\over2}$ & 1.57\\[0.5ex]
  $n=1,5$ & ${\sqrt{2}+   1    \over2}$ & 1.21\\
  \hline\hline
 \end{tabular}\\
 \end{centering}
 \caption[]{\label{enhance}
            Degeneracy enhancement for different integer fillings.}
\end{table}

\subsection{Dependence on lattice-structure}

To understand the effect of frustration in terms of the hopping 
arguments that we have made so far, we have to consider more than just one 
next-neighbor hop. 
The simplest system where we encounter frustration is a triangle with hopping 
matrix elements $-t$ between neighboring sites. In the single-electron case 
we can form a bonding
state with energy $E_{\rm min}=-2\,t,\,$ but because of frustration we
cannot form an anti-bonding state. Instead the maximum eigenenergy is
$E_{\rm max}=t.\;$ Hence frustration leads to an asymmetric 'band' of
width $W=3\,t.$ 

\begin{figure}
 \centerline{\resizebox{8cm}{!}{\mbox{
   \begin{picture}(490,250)(-196,-100)
    \thicklines
    \newcommand{\up}{$\uparrow$}
    \newcommand{\dn}{$\downarrow$}
    \newcommand{\ur}{{$\uparrow$}}
    \newcommand{\dr}{{$\downarrow$}}
    \newcommand{\levs}[3]{
     \begin{picture}(120,80)(-60,-40)
      \put(-43,-18){\line(1,0){40}}
      \put(-43,-18){\raisebox{-1.00ex}{\makebox[40pt]{\huge #1}}}
      \put( 14,-18){\line(1,0){40}}
      \put( 14,-18){\raisebox{-1.00ex}{\makebox[40pt]{\huge #2}}}
      \put(-14, 18){\line(1,0){40}}
      \put(-14, 18){\raisebox{-1.00ex}{\makebox[40pt]{\huge #3}}}
     \end{picture}
    }
    \put(-200,   0){\levs{\up\dr}{\up   }{\dn   }}
    \put( -73,-100){\levs{\up   }{\ur\dn}{\dn   }}
    \put(  73,-100){\levs{\up   }{\dn   }{\up\dr}}
    \put( 200,   0){\levs{\ur\dn}{\dn   }{\up   }}
    \put(  73, 100){\levs{\dn   }{\up\dr}{\up   }}
    \put( -73, 100){\levs{\dn   }{\up   }{\ur\dn}}
   \end{picture}
  }}}
 \caption[]{\label{triangle}
           Basis states for four electrons on a triangle in the limit of
           large Hubbard interaction $U$. They are generated e.g.\ by hopping
           the extra charge clockwise around the triangle. After three hops
           the extra charge has returned to the original site. It takes,
           however, an extra lap to restore the spins to the original 
           configuration.}
\end{figure}
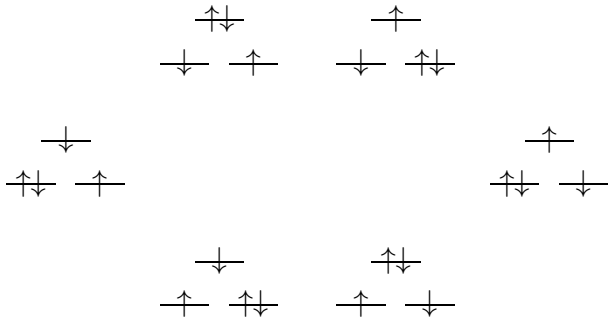

In the many-body case the situation is different. Like in the degeneracy
argument, we look at the hopping of an extra electron against a (frustrated) 
antiferromagnetic background in the large-$U$ limit. For simplicity we assume
a non-degenerate system, i.e.\ there is one electron per site on the triangle,
plus the extra electron. In this case we have to move the extra charge 
{\em twice} around the triangle to come back to the many-body state we started 
from (cf.\ Fig.\ \ref{triangle}). Thus in the large-$U$ limit the many-body 
problem is an eigenvalue problem of a $6\times6$ matrix with extreme 
eigenvalues $\pm2\,t$. In the degeneracy argument we have assumed that the 
kinetic energy of the extra charge is given by $W/2$. On the triangle, we 
find, however, that the hopping energy is by a factor of $4/3$ larger than that.
This suggests that for frustrated systems the single electron band-width $W$ 
in (\ref{Egap}) should be multiplied by a prefactor larger than one. We 
therefore expect that frustration alone, already without degeneracy, shifts the 
Mott transition to larger $U$.

To analyze the effect of frustration on the Mott transition we have determined
the critical $U$ for a hypothetical doped Fullerene A$_4$C$_{60}$ with body 
centered tetragonal (bct) structure, a lattice without frustration, having 
the same band-width ($W=0.6\,eV$) as the fcc-Fullerides, shown
in Fig.~\ref{Mott}. For $U=1.3\,eV$, we find a gap $E_g\approx0.6\,eV$ for the 
Fulleride with bct structure, while the frustrated fcc compound still is
metallic $E_g=0$. This difference is entirely due to the lattice-structure. 
Using realistic parameters for K$_4$C$_{60}$ \cite{A4C60} that crystallizes
in a bct structure we find a Mott insulator with gap $E_g\approx 0.7\,eV$, 
which is in line with experimental findings: $E_g=0.5\pm0.1\,eV$ \cite{Knupfer}.

\section{Conclusion}

We have seen that, due to more efficient hopping, orbital degeneracy increases 
the critical $U$ at which the Mott transition takes place. This puts 
the integer-doped Fullerenes close to a Mott transition. Whether they are on
the metallic or insulating side depends on the filling of the band and the
lattice-structure: Since the degeneracy enhancement is most efficient for a 
half-filled band, systems doped away from half-filling tend to be more 
insulating.  The effect of frustration, on the other hand, is to make the 
system more metallic.

\begin{ack}
This work has been supported by the Alexander-von-Humboldt-Stiftung under the
Feodor-Lynen-Program and the Max-Planck-Forschungspreis, and by the Department
of Energy, grant DEFG 02-96ER45439.
\end{ack}

\end{document}